\documentclass[fleqn,preprint,showpacs,preprintnumbers,amsmath,amssymb]{revtex4}
\usepackage{graphicx}
\usepackage{dcolumn}
\usepackage{bm}

\begin{document}

\title{Generalized matching criterion for electrostatic ion solitary propagations in quasineutral magnetized plasmas}
\author{M. Akbari-Moghanjoughi}
\affiliation{Azarbaijan University of
Tarbiat Moallem, Faculty of Sciences,
Department of Physics, 51745-406, Tabriz, Iran}

\date{\today}
\begin{abstract}
Based on the magnetohydrodynamics (MHD) model, an exact arbitrary-amplitude general solution is presented for oblique propagation of solitary excitations in two- and three-component quasineutral magnetoplasmas, adopting the standard pseudopotential approach. It is revealed that the necessary matching criterion of existence of such oblique nonlinear propagations in two and three-fluid magnetoplasmas share global features. These features are examined for the cases of electron-ion and electron-positron-ion magnetoplasmas with diverse equations of state. This study also reveals that for electron-ion magnetoplasmas with plasma-frequencies larger than the cyclotron-frequency ($B_0<0.137\sqrt{n_0}$) a critical-angle of $\beta_{cr}=\arccos{\left[B_0 /(0.137 \sqrt{n_0})\right]}$ exists, at which propagation of solitary excitation is not possible. Coriolis effect on allowed soliton matching condition in rotating magnetoplasmas is also considered as an extension to this work. Current investigation can have important implications for nonlinear wave dynamics in astrophysical as well as laboratory magnetoplasmas.
\end{abstract}

\keywords{Sagdeev potential, Electron-ion magnetoplasmas, Quasi-neutral plasmas}

\pacs{52.30.Ex, 52.35.-g, 52.35.Fp, 52.35.Mw}
\maketitle

\section{Introduction}

The collective nonlinear excitations are of fundamental properties of ionized environments, which occur in periodic and solitary forms. The first experimental evidence of the ion-acoustic solitary excitations was realized by H. Ikezi, et.al, in 1970 \cite{ikezi}. Also, Sagdeev pseudo-potential method \cite{sagdeev}, is one of the oldest \cite{vedenov} and remarkably simple but effective approaches for investigation of the nonlinear propagations in plasmas, which is analogous to the classical problem of a particle trapped in a potential. During the past few decades, there has been many investigations on the nonlinear wave propagations in diverse magnetized and unmagnetized plasma kinds \cite{popel, nejoh, mahmood1, mahmood2,
mahmood3, abdolsalam}, using Sagdeev approach. However, even in a simple two-fluid plasma interesting wave phenomenon occur due to a large mass difference between electrons and ions. Recently, it has been shown \cite{akbari1} that in a simple electron-ion quantum plasma the criterion for the existence of solitary excitations is fundamentally governed by the system dimensionality and the ion degrees of freedom \textbf{(DoF)}. On the other hand, it has been remarked \cite{akbari2} that the ion-temperature can have crucial importance on nonlinear wave dynamics in degenerate electron-positron-ion plasmas. Historically, the nonlinear ion-acoustic wave theory has been developed to include a finite ion-temperature effects \cite{tappert, tagare} because of quantitative discrepancies between theory and experiment.

One of the greatest challenges of plasma research is to understand the nonlinear processes in astrophysical environments. For instance, the origin of high-energy superthermal (non-Maxwellian) charged particles, observed in solar wind, magnetosphere, interstellar medium and auroral zone \cite{mendis, lazar}, is one of the unsolved problems in the field of space and astrophysical plasmas. Many dense astrophysical objects such as white dwarfs, usually accompanied by strong magnetic fields, are considered as completely degenerate alike ordinary metals \cite{chandra1}, in which the inter-electron distances are much lower than the characteristic de Broglie thermal wavelength, $h/(2\pi m k_B T)^{1/2}$ \cite{Bonitz}. The electron degeneracy state, such as that encountered in normal metals is caused by Pauli exclusion rule \cite{landau}. The state of plasma degeneracy has been recently studied using the remarkably simple and effective Thomas-Fermi approximation, where, it has been shown \cite{akbari3, akbari7, akbari8} that electron degeneracy may lead to distinct features in nonlinear solitary interactions and propagations. For ultradense plasmas with electron densities as high as $10^{18} cm^{-3}$ (such as in white dwarfs), where the exchange effects may be dominant, the Thomas-Fermi approximation has to be extended to Thomas-Fermi-Dirac approximation \cite{levine}. Investigations confirm \cite{akbari4} that, the relativistically degenerate Fermi-Dirac dusty plasma, unlike the simple Thomas-Fermi counterpart, accommodates a wide variety of nonlinear excitations. In some massive white dwarfs, the catastrophic gravitational pressure may lead to the extreme relativistic degeneracy state, consequently, giving rise to the total collapse of the star \cite{chandra1}. The thermodynamical features of a superdense degenerate electron gas has been studied in Refs. \cite{chandra2, kothary} in both nonrelativistic and ultrarelativistic limits, employing the Fermi-Dirac quantum statistics.

For fusion-like astrophysical compact object such as a white-dwarf the partial or finite-temperature degeneracy assumption may be more relevant compared to the zero-temperature degeneracy model. Unfortunately, the effect of finite-temperature electron degeneracy has not been paid sufficient attention in the study of nonlinear wave dynamics in superdense plasmas, whereas, it is the most relevant situation for astrophysical cases. In this work we examine the most general criterion for the existence of the solitary ion excitations in magnetized plasmas. We also demonstrate the relevance of this study for a many instances of electron-ion and electron-positron-ion magnetoplasmas, with electron distributions ranging from classical Maxwell to relativistic Fermi-Dirac electrons/positrons. The presentation of the article is as follows. The basic normalized plasma equations are introduced in Sec. \ref{equations}. A general nonlinear arbitrary-amplitude solution is derived in Sec. \ref{Sagdeev}. Applications to electron-ion magnetoplasmas are presented in Sec. \ref{discussion1}. The study is extended to electron-positron-ion magnetoplasmas in Sec. \ref{discussion2}. The effect of Coriolis force is considered in Sec. \ref{discussion3}. Finally, conclusions are drawn in section \ref{conclusion}.

\section{Basic Magnetohydrodynamic Model}\label{equations}

We consider a collisionless quasineutral uniformly magnetized plasma consisting of positively and negatively charged species. The magnetic field, ${\bf{B_0}}$, is directed along the $x$ axis. The basic conventional magnetohydrodynamics equations, then, take the following form
\begin{equation}\label{dimensional}
\begin{array}{l}
\frac{{\partial {n_\pm}}}{{\partial t}} + \nabla \cdot({n_\pm }{{\bf{u}}_\pm}) = 0,\hspace{3mm}{{\bf{u}}_\pm} = {\bf{\hat x}}{u_{x\pm}} + {\bf{\hat y}}{u_{y\pm}} + {\bf{\hat z}}{u_{z\pm}}, \\
\frac{{\partial {{\bf{u}}_ + }}}{{\partial t}} + ({{\bf{u}}_ + }\cdot\nabla ){{\bf{u}}_ + } + \frac{{e}}{{{m_+}}}\nabla \phi  + \frac{1}{{{m_+}{n_+}}}\nabla {P_+(n_+) } - {\omega _{c+}}({{\bf{u}}_ + } \times {\bf{\hat x}}) = 0, \\
\frac{{{m_ - }}}{{{m_ + }}}\left( {\frac{{\partial {{\bf{u}}_ - }}}{{\partial t}} + ({{\bf{u}}_ - }\cdot\nabla ){{\bf{u}}_ - } + {\omega _{c-}}({{\bf{u}}_ - } \times {\bf{\hat x}})} \right) + \frac{1}{{{m_ + }{n_ - }}}\nabla {P_-(n_-) } = \frac{e}{{{m_ + }}}\nabla \phi,\\
n_{-} \approx n_{+},
\end{array}
\end{equation}
where, $u_\pm$, $m_\pm$ and $\omega_{c\pm}=eB_0/m_\pm$ are the positive/negative charged species velocity, mass and cyclotron frequencies, respectively. In order to obtain a dimensionless set of equations we use general scaling defined below
\begin{equation}\label{nm}
\nabla \to \frac{1}{\lambda_+}\bar \nabla,\hspace{2mm}t \to \frac{{\bar t}}{{{\omega _{p+}}}},\hspace{2mm}{n_{\pm}} \to {n_0}{\bar n_{\pm}}, \hspace{2mm}{\bf{u_\pm}} \to {c_+}{\bf{\bar u_\pm}},\hspace{2mm}\phi  \to \frac{\epsilon}{e}\bar \phi,\hspace{2mm}P_{\pm} \to \epsilon \bar P_{\pm},
\end{equation}
where, $\omega_{p+}=\sqrt{4\pi e^2 n_0/m_+}$, ${\lambda_+} = c_+/\omega_{p+}$ and $c_+=\sqrt{\epsilon/m_+}$ are the characteristic plasma frequency, ion gyroradius and sound-speed the values of which along with the parameter $\epsilon$ will be defined later based on the charge distribution. Using the quasi-neutrality condition and assuming $m_+\gg m_-$, the normalized set of equations (i.e. ignoring the bar notation) read as
\begin{equation}\label{normal}
\begin{array}{l}
\frac{{\partial n}}{{\partial t}} + \nabla \cdot(n{\bf{u_+}}) = 0, \\
\frac{{\partial {\bf{u_+}}}}{{\partial t}} + ({\bf{u_+}}\cdot\nabla ){\bf{u_+}} + \nabla \phi + \frac{{\nabla {P_+(n)}}}{n} - \bar\omega({\bf{u_+}} \times {\bf{\hat x}}) = 0, \\
\nabla \phi  = \frac{{\nabla {P_-(n)}}}{n}, \\
\end{array}
\end{equation}
where, $\bar\omega=\omega_{c+}/\omega_{p+}$ is the normalized magnetic field strength. Taking $\bf{u_+}=\bf{u}$, we rewrite
\begin{equation}\label{normal2}
\begin{array}{l}
\frac{{\partial n}}{{\partial t}} + \nabla \cdot(n{\bf{u}}) = 0, \\
\frac{{\partial {\bf{u}}}}{{\partial t}} + ({\bf{u}}\cdot\nabla ){\bf{u}} + \frac{{\nabla P(n)}}{n} - \bar\omega({\bf{u}} \times {\bf{\hat x}}) = 0,\hspace{2mm}P(n) = {P_+(n)} + {P_-(n)}. \\
\end{array}
\end{equation}
\textbf{Now, without loss of generality, lets consider only two-dimensional perturbations in $x$-$y$ plane for simplicity. Hence, the simplified set of equations to be solved are as follows}
\begin{equation}\label{scalar}
\begin{array}{l}
{\partial _t}n + {\partial _x}(n{u_x}) + {\partial _y}(n{u_y}) = 0, \\
{\partial _t}{u_x} + \left( {{u_x}{\partial _x} + {u_y}{\partial _y}} \right){u_x} + \frac{{{\partial _x}P(n)}}{n} = 0, \\
{\partial _t}{u_y} + \left( {{u_x}{\partial _x} + {u_y}{\partial _y}} \right){u_y} + \frac{{{\partial _y}P(n)}}{n} - \bar\omega{u_z} = 0, \\
{\partial _t}{u_z} + \left( {{u_x}{\partial _x} + {u_y}{\partial _y}} \right){u_z} + \bar\omega{u_y} = 0. \\
\end{array}
\end{equation}
In the next section we will solve Eqs. (\ref{scalar}) in the most general form without using specific equation of state for charged particles.

\section{General Arbitrary-amplitude Solution}\label{Sagdeev}

Here, we seek for the most general appropriate Sagdeev pseudo-potential which describes the possibility of nonlinear electrostatic excitations in the magnetoplasma given by Eq. (\ref{normal}). Being interested in stationary wave solutions moving at constant velocity, we reduce Eq. (\ref{scalar}) to stationary frame reference by changing into the new coordinate $\xi=k_\parallel x+k_\perp y-M t$, where, $M=V/c_+$ is the normalized matching soliton-speed and $k_\parallel^{2} + k_\perp^{2}=1$, where, $\beta=\arccos{k_\parallel}$ is the angle between soliton propagation direction with that of the magnetic field. By a trivial integration with appropriate boundary conditions ($\mathop {\lim }\limits_{u_x,u_y  \to 0} n = 1$), the reduced form of Eqs. (\ref{scalar}) become
\begin{equation}\label{red}
\begin{array}{l}
\left( { - M + {k_\parallel}{u_x} + {k_\perp}{u_y}} \right) = - M/n, \\
\left( { - M + {k_\parallel}{u_x} + {k_\perp}{u_y}} \right){d_\xi }{u_x} + \frac{{{k_\parallel}}}{n}{d_\xi }P(n) = 0, \\
\left( { - M + {k_\parallel}{u_x} + {k_\perp}{u_y}} \right){d_\xi }{u_y} + \frac{{{k_\perp}}}{n}{d_\xi }P(n) - \bar\omega{u_z} = 0, \\
\left( { - M + {k_\parallel}{u_x} + {k_\perp}{u_y}} \right){d_\xi }{u_z} + \bar\omega{u_y} = 0. \\
\end{array}
\end{equation}
Combination of the equation set (\ref{red}) gives rise to the following general differential relation
\begin{equation}\label{sol}
\frac{d}{{d\xi }}\left\{ {\frac{1}{n}\left[ {\frac{{{d^2}}}{{d{\xi ^2}}}\left( {\frac{{{M^2}}}{{2{n^2}{{\bar \omega }^2}}} + \Psi (n)} \right) + 1} \right]} \right\} + \frac{{nk_{\parallel}^2}}{{{M^2}}}\frac{{d\Psi (n)}}{{d\xi }} = 0,\hspace{3mm}\Psi (n) = \int_1^n {\frac{{{d_n}P(n)}}{n}dn},
\end{equation}
where, $\Psi (n)$ is the generalized effective potential due to the total plasma pressure $P(n)$. After integration with appropriate boundary conditions, we obtain
\begin{equation}\label{sol2}
\frac{{{d^2}}}{{d{\xi ^2}}}\left[ {\frac{{{M^2}}}{{2{n^2}\bar\omega^2}} + \Psi (n)} \right] + \frac{{nk_{\parallel}^2}}{{{M^2}}}\int_1^n {n{d_n}\Psi (n)dn} - n + 1=0 .
\end{equation}
Finally, a trivial algebraic manipulation assuming the same boundary conditions, leads to the well-known energy integral of the form
\begin{equation}\label{energy}
\frac{1}{2}{\left( {\frac{{dn}}{{d\xi }}} \right)^2} + U(n) = 0, \\
\end{equation}
with the desired pseudo-potential, given below
\begin{equation}\label{pseudo}
\begin{array}{l}
U(n) = \\
{\left[ {{d_n}\Psi (n) - \frac{{{M^2}}}{{{n^3}{{\bar \omega }^2}}}} \right]^{ - 2}}\int_1^n {\left\{ {\left[ {{d_{n'}}\Psi (n') - \frac{{{M^2}}}{{{{n'}^3}{{\bar \omega }^2}}}} \right]\left[ {1 - n' + \frac{{n'k_{\parallel}^2}}{{{M^2}}}\int_1^{n'} {n''{d_{n''}}\Psi (n'')dn''} } \right]} \right\}dn'}  \\ = {\left[ {{d_n}\Psi (n) - \frac{{{M^2}}}{{{n^3}{{\bar \omega }^2}}}} \right]^{ - 2}}\left\{ {\frac{{{M^2}}}{{2{n^2}{{\bar \omega }^2}}} - \frac{{{M^2}}}{{2{{\bar \omega }^2}}} - \frac{{M^2}}{{n{{\bar \omega }^2}}} + \frac{{M^2}}{{{{\bar \omega }^2}}} + {\int_1^n {{d_{n'}}\Psi (n')dn'} }} \right. - \int_1^n {n'{d_{n'}}\Psi (n')dn'} \\ \left. { - \frac{{k_{\parallel}^2}}{{{{\bar \omega }^2}}}\int_1^n {\left[ {{{n'}^{ - 2}}\int_1^{n'} {n''{d_{n''}}\Psi (n'')dn''} } \right]} dn' + \frac{{k_{\parallel}^2}}{{{M^2}}}\int_1^n \left[{n'{d_{n'}}\Psi (n')} {\int_1^{n'} {n''{d_{n''}}\Psi (n'')dn''} } \right]dn'} \right\}, \\
\end{array}
\end{equation}
which, after some algebra, results in a generalized pseudo-potential of the form
\begin{equation}\label{pes}
\begin{array}{l}
U(n) = {\left[ {\frac{{{d_n}P(n)}}{n} - \frac{{{M^2}}}{{{n^3}{{\bar \omega }^2}}}} \right]^{ - 2}}\left\{ {\frac{{{M^2}}}{{2{{\bar \omega }^2}}}\left( {\frac{1}{{{n^2}}} - 1} \right) - \frac{{{M^2}}}{{{{\bar \omega }^2}}}\left( {\frac{1}{n} - 1} \right) + \int_1^n {\frac{{d_{n'}P(n')}}{n'}dn'} } - P(n) + \right. \\ P(n=1) - \frac{{k_{\parallel}^2}}{{{{\bar \omega }^2}}}\left. {\left[ {\int_1^n {\frac{{P(n')}}{{{n'^2}}}dn'}  + P(n=1)\left( {\frac{1}{n} - 1} \right)} \right] + \frac{{k_{\parallel}^2}}{{2{M^2}}}{{\left[ {P(n) - P(n=1)} \right]}^2}} \right\}. \\
\end{array}
\end{equation}
The possibility of solitary excitation relies on three conditions to satisfy, simultaneously, namely
\begin{equation}\label{conditions}
{\left. {U(n)} \right|_{n = 1}} = {\left. {\frac{{dU(n)}}{{dn}}} \right|_{n = 1}} = 0,{\left. {\frac{{{d^2}U(n)}}{{d{n^2}}}} \right|_{n = 1}} < 0.
\end{equation}
It is easily observed that the potential, Eq. (\ref{pes}), and its first derivative vanish at $n=1$, as required by the first two possibility conditions. Moreover, the direct evaluation of the second derivative of the Sagdeev potential (Eq. (\ref{pes})) at $n=1$, leads to the following generalized condition
\begin{equation}\label{con2}
{\left. {\frac{{{d^2}U(n)}}{{d{n^2}}}} \right|_{n = 1}} = \frac{{{{\bar \omega }^2}}}{{{M^2}}}\left[ {\frac{{{M^2}/{\Delta ^2} - k_\parallel^2}}{{{M^2}/{\Delta ^2} - {{\bar \omega }^2}}}} \right],\hspace{3mm}{\Delta ^2} = {\left. {{d_n}P(n)} \right|_{n = 1}}. \\
\end{equation}
It is clearly observed that, for the existence of oblique solitary propagations, the matching-speed should satisfy the following general inequality
\begin{equation}\label{con3}
\left\{ {\begin{array}{*{20}{c}}
{{k_\parallel} < \frac{M}{\Delta} < \bar \omega ;} & {\bar \omega  > {k_\parallel}}  \\
{\bar \omega  < \frac{M}{\Delta} < {k_\parallel};} & {{k_\parallel} > \bar \omega }  \\
\end{array}} \right\}
\end{equation}
However, some important features are to be noted here. First, it is remarked that, for $\omega_{p+}>\omega_{c+}$ ($\bar\omega<1$ i.e. $B_0/\sqrt{n_0}<0.137$) a critical-angle, $\beta_{cr}=\arccos{\bar\omega}\approx \arccos{\left[B_0 /(0.137 \sqrt{n_0})\right]}$, exists at which the propagation of solitary excitation is not possible. This can be of some importance for nonlinear wave dynamics in astrophysical rotating magnetoplasmas (rotation effects is considered in Sec. \ref{discussion3}). The effect of magnetoplasma rotation on kinetic Alfv\'{e}n solitary propagations has been considered in Ref. \cite{mofiz}. It is believed that strong magnetic field as high as $10^8(T)$ and number density up to $10^{20} (m^{-3})$ for the relativistic electrons and positrons can exist in the vicinity of pulsar magnetic poles \cite{krap}. Second, it is revealed that, for $\bar\omega>k_\parallel$, the lower/upper soliton matching speed-limit is independent from the external magnetic field magnitude/direction and for $\bar\omega<k_\parallel$, the lower/upper soliton matching speed-limit is independent from the external magnetic field direction/magnitude. The above condition is a fairly simple but very general relation the special cases of which will be examined in Sec. \ref{discussion1}. This generalized criterion can even be easily extended to collisional magnetoplasmas similar to the treatment presented in Ref. \cite{woo}. The present general result also justifies the special cases presented in Refs. \cite{abdolsalam, yu, karabi, yashvir, sultana, ghosh, lee, ivanov, mahmood, shukla}. In Sec. \ref{discussion2}, we will show that this generalized condition is also applicable to the most general case of electron-positron-ion magnetoplasmas. However, it should be noted that, the above condition does not apply to the case of unmagnetized plasmas.

\section{Applications in Electron-Ion Magnetoplasmas}\label{discussion1}

In this section we presents some applications of the derived generalized condition for electron-ion magnetoplasmas with different electronic distributions and $\bar\omega>1$. In all the cases presented below the pressure of ions is taken as $P_i=n^\gamma k_B T_i/\epsilon$, where $\gamma=(f+2)/f$ is the adiabatic constant with $f$ being the ion \textbf{DoF}. Note that the presented results are equally valid for both cases of $\gamma=1$ (isothermal-ion) and $\gamma> 1$ (adiabatic-ion) plasmas.

\subsection{Maxwell-Boltzmann Electron Distribution}

For the Maxwell-Boltzmann classical electron distribution, $n_e=e^{\phi}$, and normalized ion pressure of form $P_i=\sigma n^\gamma$, with $\sigma=T_i/T_e$ and $\epsilon=k_B T_e$ (see Eq. (\ref{nm})), we derive $k_\parallel\sqrt{\gamma\sigma+1}<M<\bar\omega\sqrt{\gamma\sigma+1}$. This relation predicts soliton-speeds of $V>c_+$ ($c_+=\sqrt{k_B T_e/m_i}$) for the normalized ion-temperatures of $\sigma>(\bar\omega^{-2}-1)\gamma^{-1}$.

\subsection{Non-Maxwellian Supperthermal Electron Distribution}

For Kappa distributed super-thermal electron distribution $n_e=(1-\phi/k)^{-k+1/2}$ \cite{awady} with $k>1/2$ and the same normalized ion pressure with $\sigma=T_i/T_e$ and $\epsilon=k_B T_e$, we obtain $k_\parallel\sqrt{\gamma\sigma+2k/(2k-1)}<M<\bar\omega\sqrt{\gamma\sigma+2k/(2k-1)}$. Therefore, similar to the Boltzmann electron distribution, this gives soliton-speeds of $V>c_+$ ($c_+=\sqrt{k_B T_{e}/m_i}$) for $\sigma>(\bar\omega^{-2}-2k/(2k-1))\gamma^{-1}$. It is noticed that in the very large limit of spectral index ($k\rightarrow \infty$) the soliton matching condition is reduced to that of Maxwell-Boltzmann.

\subsection{Zero-Temperature Thomas-Fermi Electron-Gas}

For degenerate Fermi electron-gas normalized pressures of the form $P_e=2n^{5/3}/5$ and $P_i=\sigma n^\gamma$ with $\sigma=T_i/T_{Fe} \ll 1$ ($T_{Fe}$ is the electron Fermi-temperature) and $\epsilon=k_B T_{Fe}$, we have, $k_\parallel\sqrt{\gamma\sigma+2/3}<M<\bar\omega\sqrt{\gamma\sigma+2/3}$ which gives soliton-speeds of $V>c_+$ ($c_+=\sqrt{k_B T_{Fe}/m_i}$) for $\sigma>(\bar\omega^{-2}-2/3)\gamma^{-1}$.

\subsection{Finite-Temperature Thomas-Fermi Electron-Gas}

For finite-temperature trapped degenerate electrons with normalized density of form $n_e=(1+\phi)^{3/2}+T^2(1+\phi)^{-1/2}$ \cite{shah} and classical ion pressure $P_i=\sigma n^\gamma$ with $\sigma=T_i/T_{Fe} \ll 1$ and $\epsilon=k_B T_{Fe}$, we get, $k_\parallel\sqrt{\gamma\sigma+2/(3-T^2)}<M<\bar\omega\sqrt{\gamma\sigma+2/(3-T^2)}$. Note that for the limit $T\rightarrow 0$ (zero-temperature Fermi electron-gas) the results given for the zero-temperature Thomas-Fermi electron distribution is deduced. Note also that, the increase in plasma temperature, $T$, leads to widening and shift of the allowed soliton-speed range to higher values. It is clearly remarked that there is an "unusual" upper-limit value for the plasma temperature $T_{max}\sim \sqrt{3}$ in finite-temperature case. In the quantum plasma cases presented below, the quantum Bohm-potential term has been neglected \cite{akbari4} compared to other terms in Eqs. (\ref{dimensional}) due to small electron-to-ion mass ratio.

\subsection{Zero-Temperature Fermi-Dirac Relativistically Degenerate Electron-Gas}

For relativistically degenerate electron properly normalized pressure, $P_e(\eta)=\frac{1}{8\eta_0^{3}}\left\{ {{\eta}\left( {2{\eta^{2}} - 3} \right)\sqrt {1 + {\eta^{2}}}  + 3\text{sinh}^{-1}{\eta}} \right\}$ \cite{chandra2}, where, $\eta=\eta_0 n^{1/3}$ is the relativity parameter $\eta_0=(n_{0}/N_0)^{1/3}$ (${N_0} = \frac{{8\pi m_e^{3}{c^3}}}{{3{h^3}}}\simeq 5.9 \times 10^{29} cm^{-3}$) is the relativistic degeneracy parameter \cite{akbari5} and the ion-pressure of form $P_i=\sigma n^\gamma$ with $\sigma=k_B T_i/(m_e c^2)\simeq 0$ and $\epsilon=m_e c^2$, we derive the matching condition, ${k_\parallel}\eta _0({1 + \eta _0^2})^{-1/4}/\sqrt{3} < M < \bar\omega\eta _0({1 + \eta _0^2})^{-1/4}/\sqrt{3}$. It is clearly noted that the increase in the relativistic degeneracy parameter, $\eta_0$, (increase in plasma mass-density \cite{akbari4}) leads to the widening and shift of the soliton-speed range to higher values. For typical electron density of a white dwarf $n_{0}\simeq N_0$ ($\eta_0\simeq 1$), we deduce $0.48{k_\parallel}< M < 0.48\bar\omega$.

\subsection{Finite-Temperature Fermi-Dirac Relativistically Degenerate Electron-Gas}

For finite-temperature relativistically degenerate-electron normalized pressure of the form \cite{chandra2}
\begin{equation}
\begin{array}{l}
{P_e(\eta)} = \frac{1}{8\eta_0^{3}}\left\{ {{\eta}\left( {2\eta^2 - 3} \right)\sqrt {1 + \eta^2}  + 3\ln \left[ {{\eta} + \sqrt {1 + \eta^2} } \right]} \right. \\ \left. { + 4{\pi ^2}{\theta ^2}{\eta}\sqrt {\eta^2 + 1}  + 7{\pi ^4}{\theta ^4}\eta^{ - 3}\sqrt {\eta^2 + 1} \left( {2\eta^2 - 1} \right)/15} + \ldots \right\}, \\
\end{array}
\end{equation}
where, $\theta=k_B T_e/(m_e c^2)$, $\eta=\eta_0 n^{1/3}$ is again the relativity parameter and $\eta_0=(n_{0}/N_0)^{1/3}$ (${N_0} = \frac{{8\pi m_e^{3}{c^3}}}{{3{h^3}}}\simeq 5.9 \times 10^{29} cm^{-3}$) is the relativistic degeneracy parameter and for ion pressure of form $P_i=\sigma n^\gamma$ with $\sigma=k_B T_i/(m_e c^2)\simeq 0$ and $\epsilon=m_e c^2$, the solitary excitations exist in the following relative speed-range
\begin{equation}
k_\parallel \sqrt {\frac{{40\eta _0^8 + 20{\pi ^2}\eta _0^4(1 + 2\eta _0^2){\theta ^2} + 7{\pi ^4}{\theta ^4}}}{{120\eta _0^6\sqrt {1 + \eta _0^2} }}}  < M < \bar\omega\sqrt {\frac{{40\eta _0^8 + 20{\pi ^2}\eta _0^4(1 + 2\eta _0^2){\theta ^2} + 7{\pi ^4}{\theta ^4}}}{{120\eta _0^6\sqrt {1 + \eta _0^2} }}}.
\end{equation}
It is remarked that, for $\theta=0$ (complete degeneracy limit) the result given for the zero-temperature Fermi-Dirac model is obtained. On the other hand, for the typical electron density of white dwarf, $n_{0}\simeq N_0$ ($\eta_0\simeq 1$), we get
\begin{equation}
k_\parallel \sqrt {0.23 + 3.5{\theta ^2} + 4{\theta ^4}}  < M < \bar\omega\sqrt {0.23 + 3.5{\theta ^2} + 4{\theta ^4}}.
\end{equation}
It is generally concluded that the increase in the normalized electron-temperature, $\theta$, leads to the broadening and shift of the allowed soliton speed-range to higher values.

\section{Extension to Electron-Positron-Ion Magnetoplasmas}\label{discussion2}

Following the methodology previously used, we rewrite Eq. (\ref{normal}) as
\begin{equation}\label{normalp}
\begin{array}{l}
\frac{{\partial n_i}}{{\partial t}} + \nabla \cdot(n_i{{\bf{u}}_i}) = 0, \\
\frac{{\partial {{\bf{u}}_i}}}{{\partial t}} + ({{\bf{u}}_i}\cdot\nabla ){{\bf{u}}_i} + \nabla \phi  + \frac{{\nabla {P_i}({n_i})}}{n_i} - \bar \omega ({{\bf{u}}_i} \times {\bf{\hat x}}) = 0, \\
\end{array}
\end{equation}
together with
\begin{equation}\label{con}
\nabla \phi  = \frac{{\nabla {P_e}({n_e})}}{{{n_e}}},\hspace{3mm} \nabla \phi  =  - \frac{{\nabla {P_p}({n_p})}}{{{n_p}}},\hspace{3mm} {n_e} - {n_p} \approx {n_i}. \\
\end{equation}
\textbf{Again considering perturbations in $x$-$y$ plane and changing} to the new coordinate $\xi=k_\parallel x+k_\perp y-M t$ and solving Eqs. (\ref{con}) with appropriate boundary conditions $\mathop {\lim }\limits_{\xi  \to \infty } \phi = 0$ and $\mathop {\lim }\limits_{\xi  \to \infty } n_{i,e,p} = \{1,\alpha,\alpha-1\}$, where $\alpha=n_{e0}/n_{i0}\geq 1$ (the special case of $\alpha=1$ leads to the quasineutral electron-ion magnetoplasma), results in a relation of form $\phi=\phi(n_i)$. In this case the Eq. (\ref{sol}) appears as
\begin{equation}\label{sol2}
\frac{d}{{d\xi }}\left\{ {\frac{1}{{{n_i}}}\left[ {\frac{{{d^2}}}{{d{\xi ^2}}}\left( {\frac{{{M^2}}}{{2n_i^2{{\bar \omega }^2}}} + \Phi_i ({n_i})} \right) + 1} \right]} \right\} + \frac{{{n_i}k_{\parallel}^2}}{{{M^2}}}\frac{{d\Phi_i ({n_i})}}{{d\xi }} = 0,
\end{equation}
Furthermore, the Sagdeev pseudopotential reads as
\begin{equation}\label{pseudo2}
\begin{array}{l}
U({n_i}) = {\left[ {{d_{{n_i}}}\Phi_i ({n_i}) - \frac{{{M^2}}}{{{n_i}^3{{\bar \omega }^2}}}} \right]^{ - 2}}\left\{ {\frac{{{M^2}}}{{2{n_i}^2{{\bar \omega }^2}}} - \frac{{{M^2}}}{{2{{\bar \omega }^2}}} - \frac{{{M^2}}}{{{n_i}{{\bar \omega }^2}}} + \frac{{{M^2}}}{{{{\bar \omega }^2}}}} { + \frac{{k_{\parallel}^2}}{{2{M^2}}}{{\left[ {\int_1^{{n_i}} {{{n_i'}}{d_{{{n_i'}}}}\Phi_i ({{n_i'}})} d{{n_i'}}} \right]}^2}} \right. \\ \left.
+ \int_1^{{n_i}} {{d_{{{n_i'}}}}\Phi_i ({{n_i'}})d{{n_i'}}}  - \int_1^{{n_i}} {{{n_i'}}{d_{{{n_i'}}}}\Phi_i ({{n_i'}})d{{n_i'}}}  - \frac{{k_{\parallel}^2}}{{{{\bar \omega }^2}}}\int_1^{{n_i}} {\left[ {{{n_i'}}^{ - 2}\int_1^{{{n_i'}}} {{{n_i''}}{d_{{{n_i''}}}}\Phi_i ({{n_i''}})d{{n_i''}}} } \right]} d{{n_i'}} \right\},
\end{array}
\end{equation}
where
\begin{equation}\label{sol4}
\Phi_i ({n_i}) = \phi ({n_i}) + \Psi_i ({n_i}),\hspace{3mm}\Psi_i ({n_i}) = \int_1^{n_i} {\frac{{{d_{{n_i}}}{P_i}({n_i})}}{{{n_i}}}d{n_i}}.
\end{equation}
The generalized pseudopotential for electron-positron-ion magnetoplasma can be written as
\begin{equation}\label{pse2}
\begin{array}{l}
U({n_i}) = {\left[ {\left( {{d_{{n_i}}}\phi ({n_i}) + \frac{{{d_{{n_i}}}{P_i}({n_i})}}{{{n_i}}}} \right) - \frac{{{M^2}}}{{{n_i}^3{{\bar \omega }^2}}}} \right]^{ - 2}}\left\{ {\frac{{{M^2}}}{{2{n_i}^2{{\bar \omega }^2}}} - \frac{{{M^2}}}{{2{{\bar \omega }^2}}} - \frac{{{M^2}}}{{{n_i}{{\bar \omega }^2}}} + \frac{{{M^2}}}{{{{\bar \omega }^2}}}} \right. \\ { + \frac{{k_{\parallel}^2}}{{2{M^2}}}{{\left[ {\int_1^{{n_i}} {{n_i'}{d_{{n_i'}}}\phi ({n_i'})d{n_i'} + \left. {{P_i}({n_i'})} \right|_1^{{n_i}}} } \right]}^2} + \int_1^{{n_i}} {{n_i'}^{ - 1}{d_{{n_i'}}}{P_i}({n_i'})d{n_i'} + \left. {\phi ({n_i'})} \right|_1^{{n_i}}} } \\ \left.  - \int_1^{{n_i}} {{n_i'}{d_{{n_i'}}}\phi ({n_i'})d{n_i'}}  + \left. {{P_i}({n_i'})} \right|_1^{{n_i}} - \frac{{k_{\parallel}^2}}{{{{\bar \omega }^2}}}\int_1^{{n_i}} {{n_i'}^{ - 2}} \left[ {\int_1^{{n_i'}} {{n_i''}{d_{{n_i''}}}\phi ({n_i''})d{n_i''} + \left. {{P_i}({n_i''})} \right|_1^{{n_i'}}} } \right]d{n_i'} \right\}. \\
\end{array}
\end{equation}
It should be noted that, in general, obtaining a trivial bounded solution of the form $\phi=\phi(n_i)$ might not be possible. However, fortunately we will only need the quantity ${d_{{n_i}}}\phi ({n_i})$, calculation of which is feasible in most of the cases. Using the general condition, Eq. (\ref{con2}), it is clearly observed that, for $\bar\omega>1$ the following similar rule applies to soliton matching-number
\begin{equation}\label{con5}
{k_\parallel} < \frac{M}{{\sqrt {{{\left. {\left[ {{d_{{n_i}}}\phi ({n_i}) + {d_{{n_i}}}{P_i}({n_i})} \right]} \right|}_{{n_i} = 1}}} }} < \bar \omega,
\end{equation}
which is equivalent to
\begin{equation}\label{con6}
{k_\parallel} < \frac{M}{{{\Delta _i}}} < \bar \omega,\hspace{3mm}{\Delta _i} = \sqrt {{{\left. {1/{d_\phi }{n_i}(\phi )} \right|}_{\phi  = 0}} + {{\left. {{d_{{n_i}}}{P_i}({n_i})} \right|}_{{n_i} = 1}}}.
\end{equation}
Here, we show the application of Eq. (\ref{con6}) to a wide variety of electron-positron-ion magnetoplasmas. As before, we assume that $P_i(n_i)=n_i^{\gamma} k_B T_i/\epsilon$, where $\gamma=(f+2)/f$ is the adiabatic constant. Respectively, for magnetoplasmas with Maxwell-Boltzmann, Kappa, Thomas-Fermi and Fermi-Dirac electron/positron distributions, we have
\begin{equation}\label{n}
\left[{\begin{array}{*{20}{c}}
{{\rm{Boltzmann}}} & {\left\{ {\begin{array}{*{20}{c}}
{{n_e} = \alpha {e^\phi }}  \\
{{n_p} = (\alpha  - 1){e^{ - \mu \phi }}}  \\
\end{array}} \right\}} & {} & {\mu  = \frac{{{T_p}}}{{{T_e}}}}  \\
{{\rm{Kappa}}} & {\left\{ {\begin{array}{*{20}{c}}
{{n_e} = \alpha {{(1 - \phi/k )}^{ - k + 1/2}}}  \\
{{n_p} = (\alpha  - 1){{(1 + \mu \phi/k )}^{ - k + 1/2}}}  \\
\end{array}} \right\}} & {} & {\mu  = \frac{{{T_p}}}{{{T_e}}}}  \\
{{\rm{Thomas - Fermi}}} & {\left\{ {\begin{array}{*{20}{c}}
{{n_e} = \alpha {{(1 + \phi )}^{3/2}}}  \\
{{n_p} = (\alpha  - 1){{(1 - \mu \phi )}^{3/2}}}  \\
\end{array}} \right\}} & {} & {\mu  = \frac{{{T_{Fp}}}}{{{T_{Fe}}}}}  \\
{{\rm{Fermi - Dirac}}} & {\left\{ {\begin{array}{*{20}{c}}
{\phi  = \sqrt {1 + {\alpha ^{2/3}}{n_e}^{2/3}\eta _0^2}  - \sqrt {1 + {\alpha ^{4/3}}\eta _0^2} }  \\
{\phi  = \sqrt {1 + {{(\alpha  - 1)}^{4/3}}\eta _0^2}  - \sqrt {1 + {{(\alpha  - 1)}^{2/3}}{n_p}^{2/3}\eta _0^2} }  \\
\end{array}} \right\}} & {} & {\alpha  = \frac{{{n_{e0}}}}{{{n_{i0}}}}}  \\
\end{array}}\right].
\end{equation}
The normalized matching soliton-speed in this case is also bounded through inequality of the form, ${k_\parallel}< M/ \Delta_i < \bar \omega$ for $\bar \omega>{k_\parallel}$ and viceversa, where
\begin{equation}\label{con8}
\left[ {\begin{array}{*{20}{c}}
{{\rm{Boltzmann}}} & {\Delta_i  = \sqrt {\gamma {\sigma} + \frac{1}{{\left[ {\alpha  + \mu (\alpha  - 1)} \right]}}} } & {} & {{\sigma} = \frac{{{T_i}}}{{{T_e}}}}  \\
{{\rm{Kappa}}} & {\Delta_i  = \sqrt {\gamma {\sigma} + \frac{{2k}}{{(2k - 1)\left[ {\alpha  + \mu (\alpha  - 1)} \right]}}} } & {} & {{\sigma} = \frac{{{T_i}}}{{{T_e}}}}  \\
{{\rm{Thomas - Fermi}}} & {\Delta_i  = \sqrt {\gamma {\sigma} + \frac{2}{{3\left[ {\alpha  + \mu (\alpha  - 1)} \right]}}} } & {} & {{\sigma} = \frac{{{T_i}}}{{{T_{Fe}}}}}  \\
{{\rm{Fermi - Dirac}}} & {\Delta_i  = \frac{{{\eta _0}/\sqrt 3 }}{{\sqrt {{{(\alpha  - 1)}^{1/3}}\sqrt {1 + {{(\alpha  - 1)}^{4/3}}\eta _0^2}  + {\alpha ^{1/3}}\sqrt {1 + {\alpha ^{4/3}}\eta _0^2} } }}} & {} &  \eta_0=(\frac{n_{i0}}{N_0})^{1/3}   \\
\end{array}} \right].
\end{equation}
Note that in the limit $\alpha\rightarrow 1$ the results given above become identical with that for the electron-ion cases presented in Sec. \ref{discussion1}. From Eqs. (\ref{con8}), it is remarked in general that, the increase in the fractional ion-temperature, $\sigma$, or the ion \textbf{DoF}, $f$, shifts the soliton speed-range to higher values slightly broadening the soliton-speed range, while, the increase in fractional electron to ion number-density, $\alpha$, or \textbf{positron to electron} temperature ratio, $\mu$, shifts the soliton speed-range to lower values, slightly narrowing the corresponding speed-range. Furthermore, it is observed that the increase in relativistic degeneracy parameter, $\eta_0$, always shifts the normalized speed-range to higher values and broadens the allowed Much-number range. From standard definitions it is noted that, in the \emph{non-relativistic} Fermi-gas limit for degenerate electrons and positrons, we have $E_{Fj}=\frac{\hbar^2 k_{Fj}^2}{2m_j}$ ($j=e,p$), i.e., for the non-relativistic Thomas-Fermi case we have; $\mu=((\alpha-1)/\alpha)^{2/3}$. In the \emph{ultra-relativistic} Fermi-gas limit, we have $E_{Fj}=c\hbar k_{Fj}$, consequently, for the ultra-relativistic Thomas-Fermi case we have; $\mu=((\alpha-1)/\alpha)^{1/3}$. Note that for the Fermi-Dirac case $\eta_0\ll 1$ and $\eta_0\gg 1$ correspond, respectively, to the non-relativistic and ultra-relativistic degeneracy cases.

\section{Extension to Rotating Magnetoplasmas}\label{discussion3}

The Coriolis force which has negligible effects in laboratory scale, may play a dominating role in cosmic magnetoplasmas. Chandrasekhar \cite{chandra3}, based on his study on thermal instability, has pointed out that the Coriolis force can have significant effects on the stability of a viscous flow in the presence of magnetic field. In this section we extend the problem of oblique solitary propagations to include the Coriolis effect. This work can be readily accomplished, for instance, for the case of magnetoplasma with rotation frequency parallel/antiparallel to magnetic field direction by substitution of $\bar\omega$ in Eqs. (\ref{normal2}) or (\ref{normalp}) with $\Omega = \bar\omega \pm 2\omega_r/\omega_{pi}$, where, $\omega_r$ is the rotation angular frequency \cite{das} and other parameters have their usual meanings. Now, based on our general solution, we have for the soliton matching condition; ${k_\parallel}< M/ \Delta < \bar\omega \pm 2\omega_r/\omega_{pi}$ for $\bar\omega \pm 2\omega_r/\omega_{pi}>{k_\parallel}$ and viceversa, with the $\Delta$ values presented in Sec. \ref{discussion2} for different distribution functions. It is clearly remarked that the Coriolis force generally shifts the upper matching soliton-speed to higher/lower values by a factor $\delta V_{up}=2\omega_r\lambda_i\Delta$ (for $\bar\omega \pm 2\omega_r/\omega_{pi}>{k_\parallel}$), where, $\lambda_i=\sqrt{m_e c^2/(4\pi e^2 n_{i0})}$ is the ion gyroradius. Figure 1 depicts the variation of the effect of Coriolis force with the normalized plasma mass-density for the case of zero-temperature Fermi-Dirac magnetoplasma presented in Sec. \ref{discussion1}. It should be noted that the parameter $\eta_0$ is related also to the mass-density (of white dwarf, for instance) through the relation $\rho\simeq 2m_p n_{0}$ or $\rho(gr/cm^{3})=\rho_0 \eta_0^{3}$ with $\rho_0(gr/cm^{3})\simeq 1.97\times 10^6$, where, $m_p$ is the proton mass and the cases $\bar\rho(=\rho/\rho_0)\ll 1$ and $\bar\rho\gg 1$ correspond to the nonrelativistic and ultrarelativistic degeneracy limits, respectively \cite{akbari6}. The density $\rho_0$ is exactly in the range of a mass-density of a typical white dwarf (the density of typical white dwarfs can be in the range $10^{5}<\rho(gr/cm^{3})<10^{9}$). From Fig. 1, it is observed that, the Coriolis effect dominates for the case of normal plasma degeneracy ($\bar\rho\ll 1$) and becomes less important as the plasma get ultrarelativistically dense ($\bar\rho\gg 1$).

\section{Concluding Remarks}\label{conclusion}

We derived a general condition for existence of solitary oblique propagations in a two and three-fluid magnetoplasmas based on magnetohydrodynamics model using the Sagdeev potential approach. It has been revealed that the lower and upper matching soliton-speed limits are universally define by the strength and direction of the external uniform magnetic field. Effect of other plasma parameters on the matching speed range is only to shift the range to higher or lower values. This calculation can be used in the cases of a wide variety of electron-ion, electron-positron-ion and dusty magnetoplasmas and can be extended to multifluid magnetoplasmas with arbitrary electron/ion pressure-density functions and can be useful in both laboratory and astrophysical studies including rotational effects.

\newpage

\textbf{FIGURE CAPTIONS}

\bigskip

Figure-1

\bigskip

Variation in normalized upper soliton matching-speed, $\delta V_{up}/\omega_r$, with respect to the normalized plasma mass-density, $\bar\rho$, for the case of zero-temperature Fermi-Dirac relativistically degenerate plasma.

\begin{figure}
\resizebox{1\columnwidth}{!}{\includegraphics{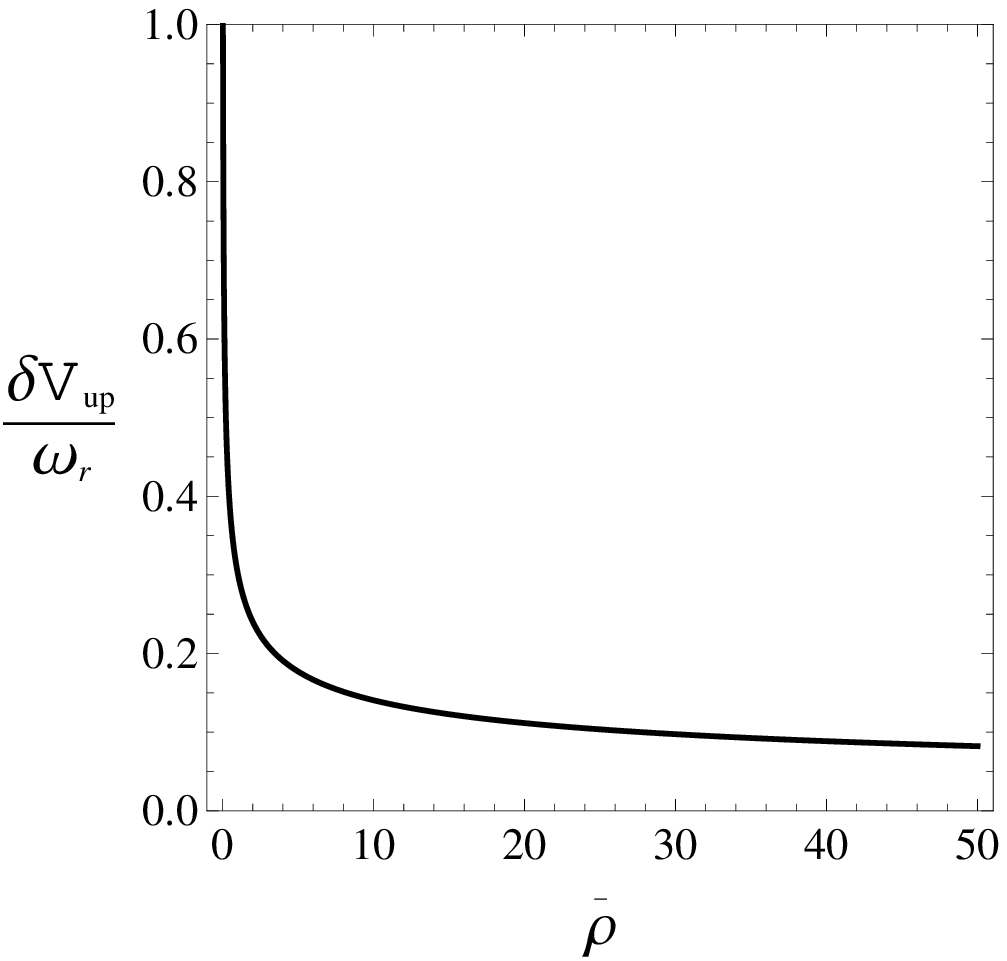}}
\caption{}
\label{fig:1}
\end{figure}

\end{document}